\documentclass[letter]{aa} 
\usepackage{graphicx}
\usepackage{txfonts}
\usepackage{natbib}
\bibpunct{(}{)}{;}{a}{}{,} 
\begin{document}
   \title{Seismic signature of envelope penetrative convection:\\ the CoRoT star HD~52265}
   \author{Y. Lebreton
          \inst{1, 2}
          \and
          M.J. Goupil\inst{3}
          }
   \institute{Observatoire de Paris, GEPI, CNRS UMR 8111, F-92195 Meudon, France
         \and
             Institut de Physique de Rennes, Universit\'e de Rennes 1, CNRS UMR 6251, F-35042 Rennes, France \\
              \email{yveline.lebreton@obspm.fr}
         \and
             Observatoire de Paris, LESIA, CNRS UMR 8109, F-92195 Meudon, France
             }
   \date{Received , 2012; accepted , 2012}

 
  \abstract
   {}
   {We aim at characterizing the inward transition from convective to radiative energy transport  at the base of 
   the convective envelope of the solar-like oscillator \object{HD~52265} recently observed 
       by the {\small CoRoT} satellite.}
   {We investigated the origin of one specific  feature found in the \object{HD~52265} frequency spectrum. We modelled the star to derive the internal structure and the oscillation frequencies that best match the observations and used a seismic indicator sensitive to the properties of the base of the envelope convection zone.}
   { The seismic indicators clearly reveal that to best represent the observed properties of \object{HD~52265}, models must include penetrative convection below the outer convective envelope. The penetrative distance is estimated to be $\sim0.95 H_P$, which corresponds to an extent over a distance representing $6.0$ per cents of the total stellar radius, significantly larger than what is found  for the Sun. The inner boundary of the extra-mixing region is found at $0.800\pm0.004\ R$ where $R=1.3\ R_\odot$ is the stellar radius.}
  {These results  contribute to the tachocline characterization in stars other than the Sun.}


 \maketitle
%

\section{Introduction}

Low-mass main-sequence (MS) stars have  convective envelopes in which 
  the chemical elements are mixed on short time scales and - at least in the deeper parts of the envelope - the energy transport by fluid elements can be treated as an adiabatic process. In the standard description, convective zones (CZ) are regions where the Schwarzschild criterion is fulfilled. Their boundaries lie at the border where the adiabatic temperature gradient 
 equals the radiative one. 
It is  expected, however, that fluid elements penetrate into the 
adjacent radiative zone due to their inertia \citep[see e.g.][]{1991A&A...252..179Z}. 
 In the Sun, the transition region (tachocline) is believed to be the site where the dynamo originates. 

Penetrative convection (PC) corresponds 
to efficient convective heat transport - and material mixing - 
by downward flows that establish a close to adiabatic temperature
 stratification below which the downward plumes 
 are no longer able to modify the temperature stratification that 
 remains close to radiative. 
The extent of penetrative convection and/or overshoot in stars cannot be derived from first principles and is 
still largely unknown.

A crucial point therefore is to find observational signatures 
of penetrative convection in low-mass MS stars. This can be achieved by means of asteroseismology.
The abrupt change of energy transport from a convective to a radiative regime
is visible in the sound speed profile and impacts
 the oscillations frequencies as well as some characteristic frequency spacings,
  which then show an oscillatory (periodic) behaviour 
  \citep{1990LNP...367..283G,1994MNRAS.268..880R}. 
Owing to the importance of the tachocline region, helioseismic studies have attempted 
to measure the extent of the PC in the Sun \citep{1993ASPC...40...60B, 1994A&A...283..247M}. 
\citet{2011MNRAS.414.1158C} have recently  found that convective envelope 
overshoot is necessary over an estimated extent of $0.37 H_P$ ($H_P$ is the pressure scale height). The high-quality solar data and wide available range of values of mode degrees $\ell$ allowed \citeauthor{2011MNRAS.414.1158C} to also show that the transition between the convective and the radiative stratification must be smooth, intermediate between a classical - ballistic - overshoot formulation and a no-overshoot one.   

Solar-like oscillations have been identified in many stars first from the 
ground, then from space by the {\small CoRoT} \citep{2002ESASP.485...17B} and 
Kepler \citep{2010cosp...38.2513K} high-precision photometry missions, leading to  an accuracy in frequency measurements of a few tenths $\mu$Hz. Theoretical studies 
of the effects on the oscillations frequencies of PC at the base of a CZ have been 
conducted using low-degree modes, the only ones expected to be detectable in stars 
\citep[see e.g.][ and references therein]{2009A&A...493..185R}. 
The period of the oscillatory component is found to be related to the location of the discontinuity inside the star and its amplitude to depend on the height of the discontinuity. One then expects that the 
amplitude of the oscillatory signal grows with an increasing extent of PC that 
causes a larger jump from the adiabatic temperature gradient to the radiative one below.
 Moreover, for a larger PC extent, the discontinuity is located deeper inside the star and the period of the oscillation is expected to be longer \citep{1993ASPC...42..173R}. 

\citet{2011A&A...530A..97B} analysed the {\small CoRoT} oscillation spectrum of the  star \object{HD~52265} (\object{HIP~33719}),
a high-metallicity G0V star hosting an exoplanet. They found  a typical 
p-mode solar-like spectrum, and identified 28 reliable low-degree p-modes of degrees $\ell=0, 1, 2$ and order $n$ in the range $14-24$ (see their Table\ 4). The frequencies $\nu_{n, \ell}$ are in the range $1500-2550\ \mu$Hz with a frequency at maximum amplitude $\nu_\mathrm{max}=2090\pm 20\mu$Hz. The error on each frequency is a few tenths of 
$\mu$Hz. Such a high quality data set enables to probe the interior structure of the star. 
Here we report on one evidence for penetrative convection below the
upper convective region of  \object{HD~52265} as indicated by its internal structure modelling.

\section{\object{HD~52265}: observations and modelling}

\subsection{Global and seismic observational constraints}

To model \object{HD~52265} we adopted the effective temperature $T_\mathrm{eff}=6120\pm110$ K and metallicity $\mathrm{[Fe/H]}=0.22\pm0.05$ dex that we derived from an average of 20 spectroscopic determinations reported since 2001. We adopted the luminosity $L=2.053\pm0.053 L_\odot$ that we derived from the Hipparcos parallax \citep[][]{2007ASSL..350.....V}, Tycho magnitude and bolometric correction calculated according to \citet{2003AJ....126..778V}. We related the ratio of heavy elements mass fraction $Z$ to hydrogen mass fraction $X$ to $\mathrm{[Fe/H]}$ 
through $\mathrm{[Fe/H]}=\log(Z/X)-\log(Z/X)_\odot$ and adopt $(Z/X)_\odot=0.0244$ from the \citet{1993oee..conf...15G} solar mixture.

\begin{figure}
      \resizebox{\hsize}{!}
	     {\includegraphics{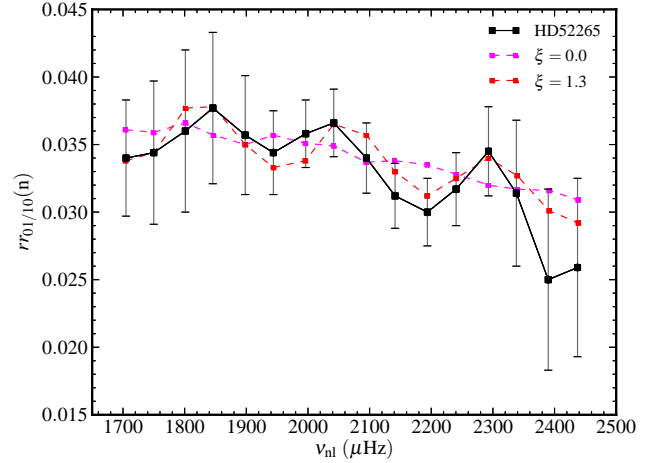}}
  \caption{
Observed ratios $rr_{01/10}(n)$ (black squares) as a function of frequency 
$\nu_{n, 0/n, 1}$ for \object{HD~52265} {compared to the $rr_{01/10}(n)$ ratios in} the 
models without PC ($\xi=0.0$, magenta) and with PC ($\xi=1.3$, red).
    }
\label{fig.rr10obs-mod}
\end{figure}

In the following we will focus on the ability of models to reproduce the values of the seismic indicators $rr_{01}(n)$ and $rr_{10}(n)$ introduced by \citet[][]{2003A&A...411..215R}, which are defined as  
\begin{equation}\label{eq.rr10}
rr_{01}(n)=dd_{01}(n)/\Delta \nu_{1}(n)  ~~~~;~~~~ rr_{10}(n)=dd_{10}(n)/\Delta \nu_{0}(n),
 \end{equation}
 where 
 \begin{eqnarray}
dd_{01}(n)&=& \frac{1}{8}(\nu_{n-1, 0}-4\nu_{n-1, 1}+6\nu_{n, 0}-4\nu_{n, 1}+\nu_{n+1, 0}) \\
dd_{10}(n)&=& -\frac{1}{8}(\nu_{n-1, 1}-4\nu_{n, 0}+6\nu_{n, l}-4\nu_{n+1, 0}+\nu_{n+1, 1}),
\end{eqnarray}
where $\Delta \nu_{\ell}(n) =\nu_{n+1,\ell}-\nu_{n,\ell}$ are 
the standard large frequency separations \citep{1980ApJS...43..469T}. 
\citeauthor[][]{2003A&A...411..215R} showed that the ratios $rr_{01}$ and  $rr_{10}$ (hereafter $rr_{01/10}$) are  sensitive to the sharp 
variation of the sound speed  in the transition region between the upper CZ and 
the radiative layers beneath \citep[see also][]{2005MNRAS.356..671O}. 
\citet[][]{2009A&A...493..185R}, using high-quality solar  data,
 confirmed that these indicators exhibit an oscillatory behaviour as 
 a function of the frequency. {From the $28$ reliable individual frequencies determined by \citet{2011A&A...530A..97B}, we calculated the $rr_{01/10}(n)$ ratios.} The variations of $rr_{01}(n)$ 
 (resp. $rr_{10}(n)$) as a function of frequency $\nu_{n, 0}$ (resp. $\nu_{n, 1}$) are 
 plotted in Fig~\ref{fig.rr10obs-mod}. \citeauthor{2011A&A...530A..97B} already noticed a periodic 
 variation of the separations $d_{01}(n)= \nu_{0, n-1}-0.5(\nu_{1, n} +\nu_{1, n-1})$ and 
 $d_{10}(n)=-\nu_{1, n}+0.5(\nu_{0, n} +\nu_{0, n+1})$  and suggested that it could be the 
 signature of the CZ boundary. The periodic variation is even clearer in the $rr_{01/10}(n)$ 
 separation, and Section~\ref{PC} is devoted to confirming the origin of this periodic signal
  and to characterizing it.

\subsection{Internal structure models and their oscillations}

We modelled \object{HD~52265} with the evolution code Cesam2k 
\citep{2008Ap&SS.316...61M} and the input physics and parameters 
described in \citet{2011arXiv1108.6153L}. 
We considered  the model for penetrative convection below 
  the convective envelope proposed by \citet[][]{1991A&A...252..179Z}. In this model the distance
   of fluid penetration into the radiative 
  zone reads $L_\mathrm{p}=(\xi/\chi_{P}) H_P$, where $\chi_P = (\partial \log \chi/\partial \log P)_\mathrm{ad}$ 
  is the adiabatic derivative with respect to pressure $P$ of the radiative 
  conductivity $\chi =16 \sigma T^3/(3 \rho \kappa)$ ($T, \rho, \kappa, \sigma$ 
  are the temperature,  density, opacity, and Boltzmann constant, respectively). 
  The free parameter $\xi$ is of the order of unity but has to be calibrated by 
  comparing stellar models to observations.
 The seismic indicators $rr_{01/10}$ that we considered here are sensitive to the change 
  in the temperature derivative hence to the transition from an unstable to a stable stratification.
 The amplitude of the periodic signal is smaller for a smoother transition 
 and cannot be detected if the transition is highly smooth. 
  We  therefore concentrated on determining  the adiabatic extent of the overshoot region. We 
    imposed accordingly that the temperature gradient in the overshooting zone
  is the adiabatic gradient.

Equilibrium models were calculated and adjusted to satisfy the constraints provided by 
 the  global parameters ($L, T_\mathrm{eff}$ and  surface $\mathrm{[Fe/H]}$) and the $28$ reliable observed frequencies of \citet{2011A&A...530A..97B}. The model frequencies were calculated with the LOSC adiabatic  oscillation code \citep{2008Ap&SS.316..149S} for the whole range of observed mode orders and degrees, and the observed and modelled seismic indicators were derived consistently. We corrected the model frequencies from the so-called near surface effects using the empirical correction proposed by \citet{2008ApJ...683L.175K}.
The frequency differences may be sensitive to surface effects, however, we recall that \citet{2003A&A...411..215R} demonstrated that the ratios of small to large 
separations are quite independent of the surface treatment and are therefore efficient probes of the interior. 

We used the Levenberg-Marquardt minimization method as described in \citet[][]{2005A&A...441..615M}  to adjust the age and mass, the initial helium and metallicity, the mixing-length parameter for convection, the convective core overshooting extent, and surface effects parameter so that the model of \object{HD~52265} fits the observations best, within the error bars. 
The goodness of fit was evaluated through minimization of the reduced $\chi^2$:
$\chi^2 = (N_\mathrm{obs}-1)^{-1} \cdot\sum_{i=1}^{N_\mathrm{obs}} 
\left( (x_\mathrm{i, mod} - x_\mathrm{i, obs}) /\sigma_\mathrm{i, obs}\right)^2
$,
\label{chi2}
where $N_\mathrm{obs}$ is the number of observational constraints considered, 
$x_\mathrm{i, mod}$ and  $x_\mathrm{i, obs}$ are the modelled and observed values of the $i^\mathrm{th}$ parameter, respectively, and $\sigma_\mathrm{i, obs}$ is the error on $x_\mathrm{i, obs}$. {We point out that the convective core overshooting extent derived from preliminary model calibrations of \object{HD~52265} is low, lower than $0.10 H_p$. Because the properties of the central region do not affect the effect we study at the transition region, the models presented here do not include core overshooting.} The detailed properties of the models will be presented in a forthcoming paper.

\begin{figure}
      \resizebox{\hsize}{!}
            {\includegraphics[]{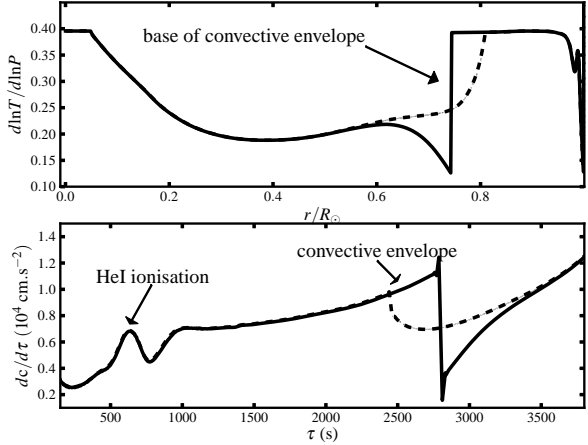}}
  \caption{Top: Temperature gradient as a function of radius in a model with penetrative convection (continuous line)  and a standard model       (dashed line) of \object{HD~52265}. Bottom: Sound speed gradient with respect to the acoustic depth $\tau(r)=\int_r^R ~dr/c_s$ for the same models.
    }
\label{fig.sharp_features}
\end{figure}

\section{Extent of penetrative convection}
\label{PC}

The best-fit models yield total {$\chi^{2}$-values in the range $2$-$3$ whatever the value of the PC and therefore cannot be distinguished on that basis. The models have an age $A\simeq$$2$ Gyr, a mass $M\simeq1.25\ M_\odot$ and a radius $R\simeq1.3\ R_\odot$ {in quite reasonable agreement with the values derived by \citet{2012A&A...543A..96E} in their recent modelling of the star}.  We now examine the impact of PC on the $rr_{01/10}(n)$ frequency ratios.

We first compared models with  $\xi=0.0$ and $1.3$. The variations of the  ratios $rr_{01/10}(n)$ with 
frequency are shown in Fig.\ref{fig.rr10obs-mod}. The amplitude of the oscillatory component is significantly smaller for $\xi=0.0$  than for $\xi={1.3}$. 
The impact of  a nonvanishing $\xi$ on the structure of the model with $\xi=1.3$ 
compared with a model without PC is illustrated in Fig.\ref{fig.sharp_features}. 
Differences appear in the  transition region below the convection zone while everywhere else, the  structures coincide.  
 The top of Fig.~\ref{fig.sharp_features} shows the behaviour of the temperature gradient below 
the CZ in the model of \object{HD~52265}. Without PC 
the CZ is located at $r_\mathrm{Sc}$, the Schwarzschild radius. The temperature gradient drops from its adiabatic value 
$0.4$ at the border of the convective envelope and takes the value of the radiative gradient.
In contrast, when PC takes place below $r_\mathrm{Sc}$ as in model with $\xi=1.3$, a discontinuity 
in the temperature gradient occurs at the bottom of the mixed region 
(including the PC extent), resulting in a nearby density discontinuity. 
This sharpens the variation of the adiabatic sound speed $c_\mathrm{s}=(\Gamma_1 P/\rho)^{1/2}$ at the base of the convective envelope in models including PC ($\Gamma_1$ is the first adiabatic index). The bottom of Fig.~\ref{fig.sharp_features}  shows the sound speed gradient as a function 
 of the acoustic depth $\tau(r)=\int_r^R ~dr/c_s$.
Rapid variations in the sound speed gradient can be seen, which correspond to the HeI ionization ($\tau\sim620$ s)
 and to the inner limit of the external CZ. At the base of the CZ,  
  the variation of  $dc_s/d\tau$ for the model 
  including PC  behaves as a near discontinuity  ($\tau\sim 2780$ s), whereas is remains continuous  
   in the standard model where only  $d^2c_s/d\tau^2$ is marginally discontinuous ($\tau\sim 2450$ s). 

To examine the ability of models to reproduce the observed $rr_{01/10}$ oscillatory trend, the difference between the signals from the observations and  models was measured with a specific $\chi^2$ quantity where the variables are $x_i= {rr_{01/10}(n)}$. Considering models with increasing $\xi$,  we found that the
variation of $\chi^2$  shows a minimum between
$\xi=1.2$ and $1.3$. The signals from models with $\xi \le 1.1$ or $\xi \ge 1.4$  lead to significantly larger $\chi^2$, excluding these models as acceptable. We therefore found a PC extent of $\xi\simeq1.25\pm 0.10$  for \object{HD~52265}. 
This translates into an overshoot distance of $d_\mathrm{ov}=0.95 \pm 0.08~H_p$  corresponding to 
  $d_\mathrm{ov}=0.060\pm0.004~R$. 

\begin{figure}
      \resizebox{\hsize}{!}
           {\includegraphics{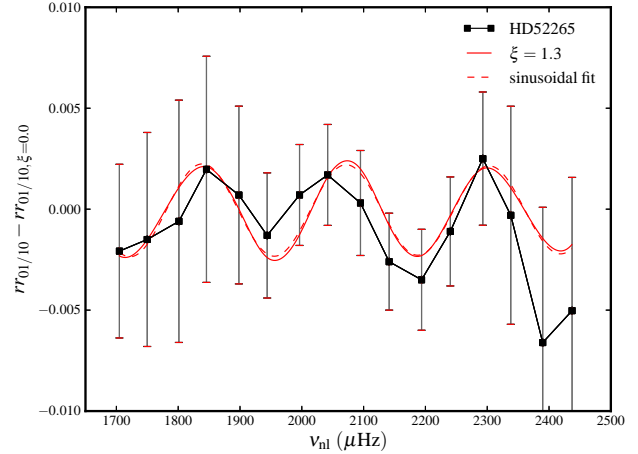} }
  \caption{
Difference of the ratios $rr_{01/10}(n)$  as a function of  frequency between the standard model (no PC,  $\xi=0.0$) and (i) the model with PC with an extent of $\xi=1.3$ (red continuous line), (ii) the observations (black). The dashed red line is a sinusoidal fit of the red curve (see text). Note that to calculate the differences, models were splined on the observed frequency grid. A detailed comparison with observations is provided in Fig.~\ref{fig.diff-rr10obs}.
    }
\label{fig.diffrr10mod}
\end{figure}

\begin{table}
\caption{Fit of the signal $S\equiv\Delta{rr_{01/10}(n)}= rr_{01/10}(n)_{S}-rr_{01/10}(n)_{\xi=0}$. 
Case~1: $ \Delta{rr_{01/10}(n)}=(a/\nu+b/\nu^2) \cos(4 \pi \nu T + 2 \phi)$+offset. 
Case~2: $\Delta{rr_{01/10}(n)}= (a/\nu) \cos(4 \pi \nu T + 2 \phi) +(b/\nu^2) \sin(4 \pi \nu T + 2 \psi)$+offset. Model fit ($M_{\xi=1.3})$ ; fit of raw observations ($O$) ; spline adjustment on observations ($Os$).
}             
\label{Tab.3}      
\centering                          
\begin{tabular}{ccccccccc}        
\hline\hline                 
signal $S$ & $T$ & $a$ & $b$ & $\phi$ & $\psi$ & offset \\
& $[\mathrm{s}]$   & $[\mu \mathrm{Hz}]$  & $[10^4 \mu\mathrm{Hz}^2]$ & $[\mathrm{rad}]$ & $[\mathrm{rad}]$ & $10^{-4} $ \\
\hline                        
Case 1& & &  &  & & &\\
\hline                        
$M_{\xi=1.3}$   & $2153$ &-8.29 & 0.75 &  -1.32 & - &-0.5\\
\hline                        
$O$  &  2187  &-34.1 & 5.76& -1.64 &-& -8.4\\
\hline                        
$Os$ &  2204  &-36.0 & 6.18 & -1.86 &-& -7.6\\
\hline                        
Case 2 &  &  &  & &  &\\
\hline                        
$M_{\xi=1.3}$   & $2182$ &-9.03 &1.03  &  -1.89 & $-1.28$ & -0.5 \\
\hline                        
$O$   & 1999  &-48.9 & 9.14 & 1.24 & 2.09& -6.3 \\
\hline                        
$Os$  &  2040  &-49.1 & 9.15 & 0.63 &1.44& -6.0\\
\hline                        
\end{tabular}
\end{table}

\section{Seismic location of the  base of the adiabatic temperature stratification}

The structure of inner regions below the transition layer is
unaffected by a change in $\xi$. It is therefore reasonable to assume 
that the long-term trend seen in Fig.\ref{fig.rr10obs-mod} for both 
signals with and without PC is the same and has its origin deeper in the star. This enables us to
remove it to isolate the oscillatory component arising 
from the base of the CZ  of model with
$\xi=1.3$.  
We interpolated the $rr_{01/10}(n)$ signals of the models with $\xi=0.0$  and $1.3$ on the same  frequency 
grid and computed their differences. 
The differences plotted in Fig.\ref{fig.diffrr10mod} are clearly generated by including PC in the  model with $\xi=1.3$.
 We obtain an almost pure sinusoid, the characteristics of which can be estimated and related to the properties
of the region of penetrative convection.
Slightly different expressions of the sinusoidal component arising from the CZ base that is expected to be visible in several seismic indicators of distant stars have been established theoretically 
 \citep[see e.g.][]{1994MNRAS.268..880R,2000MNRAS.316..165M,2004ESASP.559..313B,2005A&A...441.1079M,2006ApJ...638..440V,
 2007MNRAS.375..861H,2009A&A...493..185R}. 
 Guided by them, we assumed 
 that the model oscillation in Fig.\ref{fig.diffrr10mod}, which stems from the difference
$\Delta{rr_{01/10}(n)}= rr_{01/10}(n)_{\xi=1.3}-rr_{01/10}(n)_{\xi=0.0}$, can take two nearly
equivalent  forms expressed in the caption of Table~\ref{Tab.3}. 
\citet{2009A&A...493..185R} expressed the periodicity of the signal as $1/(2 {\mathrm{T}}_{\xi=1.3})$, which provides 
the acoustic radius  $\mathrm{T}_{\xi=1.3}=\tau_t-\tau(r)$ at the base of the CZ. $\tau(r)$ is the acoustic depth and 
 $\tau_t \equiv \tau(r=0)=1/(2\langle\Delta \nu\rangle)$, where $\langle\Delta \nu\rangle$ is
  the mean asymptotic large frequency separation.
The amplitude is a slowly varying function of $\nu$ with a term in $\nu^{-1}$
  (resp. $\nu^{-2}$), arising from a 
discontinuity in the first (resp. second) derivative of the sound speed. 
 We list in Table~\ref{Tab.3} the values of $\mathrm{T}_{\xi=1.3}$ and other parameters
of the fit of the sinusoidal signal
of Fig.\ref{fig.diffrr10mod} ($\chi^2_\mathrm{fit} =1.6~10^{-4}$). The results weakly depend on the assumed form
for $\Delta{rr_{01/10}(n)}$. We find ${\mathrm{T}}_{\xi=1.3} \simeq 2167 \pm 15$ s, which is higher than values obtained  by \citet{2000MNRAS.316..165M} because \citeauthor{2000MNRAS.316..165M}  considered less overshoot and
slightly smaller stellar masses.  
 These ${\mathrm{T}}_{\xi=1.3}$ values can now be compared with  the
theoretical value, ${\mathrm{T}}_\mathrm{ov}=\tau_t-\tau(r_\mathrm{ov})$, directly computed from the equilibrium model with $\xi=1.3$, where $r_\mathrm{ov}$ is the radius at the base of the
adiabatic region and $\tau_t= 4913$\ s for the model with $\xi={1.3}$.
The fitted value ${\mathrm{T}}_{\xi=1.3}$ falls within $\sim 35$\ s ($\sim 2\%$) of  
the resulting theoretical value ${\mathrm{T}}_\mathrm{ov}=2132$\ s, the shift ${\mathrm{T}}_{\xi=1.3}-{\mathrm{T}}_\mathrm{ov}$ being similar to that found for the Sun by \citet{2009A&A...493..185R}.

The same procedure was applied to the observed signal, which is compared to the model with $\xi=0.0$ and the sinusoidal fits are plotted in Fig.~\ref{fig.diff-rr10obs}. 
The results of the fits  ($\chi^2_\mathrm{fit} =10^{-3}$) 
are listed in Table~\ref{Tab.3}.   
The period $T_\mathrm{obs}=2100\pm 100$ s  agrees quite well  with that derived  for model $\xi=1.3$.  
We conclude that 
the base of the  adiabatically stratified region  of \object{HD~52265} is located within $2\%$ 
of the radius $r_\mathrm{ov}/R= 0.800\pm0.004$.

\begin{figure}
      \resizebox{\hsize}{!}
            {\includegraphics{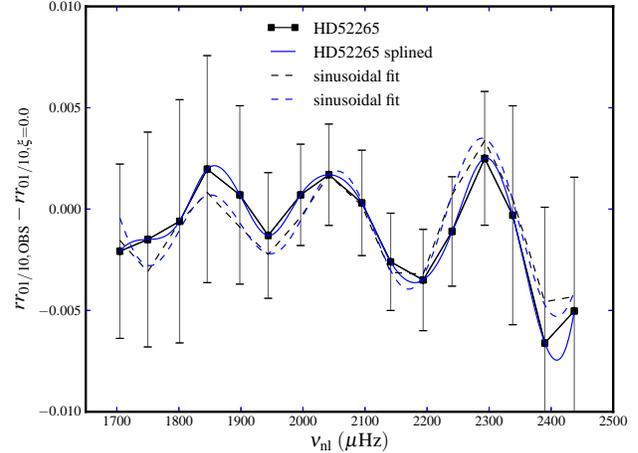}}
  \caption{
Difference of the ratios $rr_{01/10}(n)$  
as a function of  frequency between the standard model (no PC,  $\xi=0.0$) 
and the raw observations (black) or a spline fit of observations (continuous blue line). The dashed lines are the fits of the raw and 
splined data (see Table~\ref{Tab.3}).}
\label{fig.diff-rr10obs}
\end{figure}

\section{Conclusions}

We have detected  an oscillatory signal with a significant amplitude in the variation 
of the separations $rr_{01/10}(n)$ with frequency for the
{\small CoRoT} star \object{HD~52265}. 
A comparison with the same signal arising from appropriate stellar models 
shows that it cannot
be reproduced unless penetrative convection is included at the base of the outer convective envelope. 
This is the first time that such a feature  is firmly detected in a star other than the Sun.  
A best fit of the signal provides a measure of the extent of the mixed region below the CZ, in terms of the 
proxy $d_\mathrm{ov}=0.95\pm0.08~H_p$.  

The  periodic signal for \object{HD~52265} is more pronounced than for the Sun  \citep{
2011MNRAS.414.1158C}, with a longer period, and so is the measured extent of PC in
terms of $H_p$ and normalized radius.  
We point out that \object{HD~52265} is similar to the Sun in all aspects except for the higher metallicity. Therefore, to understand
 the amplitude difference between the two stars, it is important to investigate the impact of metallicity on the structure  and dynamics of
 the tachocline. {Progress is expected in a near future when seismic missions will provide similar observations
for many stars spanning a wide metallicity range.}

With the above results, 
we provided an additional clue that the physical description of the turbulent convection must be improved in stars, particularly at interfaces with radiative
regions. Future modelling of the dynamics in the region of the tachocline will have to comply with
our findings as well as with the results previously obtained for the Sun. 


\bibliographystyle{aa} 
\bibliography{yl} 

\end{document}